\documentclass[prb,twocolumn, showpaces,preprintnumbers,amsmath,amssymb,superscriptaddress]{revtex4-1}
\usepackage{graphicx}
\usepackage{dcolumn}
\usepackage{bm}
\usepackage{color}
\usepackage{amssymb}
\usepackage{amsmath}
\usepackage{epstopdf}
\usepackage{bbding}
\usepackage[mathlines]{lineno}
\usepackage[dvipdfm, pdfstartview=FitH, CJKbookmarks=true, bookmarksnumbered=true, bookmarksopen=true, colorlinks, pdfborder=001, linkcolor=black, anchorcolor=blue, citecolor=blue]{hyperref}
\usepackage{lastpage}
\usepackage{fancyhdr}
\begin{document}
\title{Extraordinary Hall resistance and unconventional magnetoresistance in Pt/LaCoO$_3$ hybrids}
\author{T. Shang}
\affiliation{Key Laboratory of Magnetic Materials and Devices, Ningbo Institute of Material Technology and Engineering, Chinese Academy of Sciences, Ningbo, Zhejiang 315201, China}
\affiliation{Laboratory of Magnetic Materials and Application Technology, Ningbo Institute of Material Technology and Engineering, Chinese Academy of Sciences, Ningbo 315201, China}
\author{Q. F. Zhan}
\email{zhanqf@nimte.ac.cn}
\affiliation{Key Laboratory of Magnetic Materials and Devices, Ningbo Institute of Material Technology and Engineering, Chinese Academy of Sciences, Ningbo, Zhejiang 315201, China}
\affiliation{Laboratory of Magnetic Materials and Application Technology, Ningbo Institute of Material Technology and Engineering, Chinese Academy of Sciences, Ningbo 315201, China}
\author{H. L. Yang}
\affiliation{Key Laboratory of Magnetic Materials and Devices, Ningbo Institute of Material Technology and Engineering, Chinese Academy of Sciences, Ningbo, Zhejiang 315201, China}
\affiliation{Laboratory of Magnetic Materials and Application Technology, Ningbo Institute of Material Technology and Engineering, Chinese Academy of Sciences, Ningbo 315201, China}
\author{Z. H. Zuo}
\affiliation{Key Laboratory of Magnetic Materials and Devices, Ningbo Institute of Material Technology and Engineering, Chinese Academy of Sciences, Ningbo, Zhejiang 315201, China}
\affiliation{Laboratory of Magnetic Materials and Application Technology, Ningbo Institute of Material Technology and Engineering, Chinese Academy of Sciences, Ningbo 315201, China}
\author{Y. L. Xie}
\affiliation{Key Laboratory of Magnetic Materials and Devices, Ningbo Institute of Material Technology and Engineering, Chinese Academy of Sciences, Ningbo, Zhejiang 315201, China}
\affiliation{Laboratory of Magnetic Materials and Application Technology, Ningbo Institute of Material Technology and Engineering, Chinese Academy of Sciences, Ningbo 315201, China}
\author{Y. Zhang}
\affiliation{Key Laboratory of Magnetic Materials and Devices, Ningbo Institute of Material Technology and Engineering, Chinese Academy of Sciences, Ningbo, Zhejiang 315201, China}
\affiliation{Laboratory of Magnetic Materials and Application Technology, Ningbo Institute of Material Technology and Engineering, Chinese Academy of Sciences, Ningbo 315201, China}
\author{L. P. Liu}
\affiliation{Key Laboratory of Magnetic Materials and Devices, Ningbo Institute of Material Technology and Engineering, Chinese Academy of Sciences, Ningbo, Zhejiang 315201, China}
\affiliation{Laboratory of Magnetic Materials and Application Technology, Ningbo Institute of Material Technology and Engineering, Chinese Academy of Sciences, Ningbo 315201, China}
\author{B. M. Wang}
\affiliation{Key Laboratory of Magnetic Materials and Devices, Ningbo Institute of Material Technology and Engineering, Chinese Academy of Sciences, Ningbo, Zhejiang 315201, China}
\affiliation{Laboratory of Magnetic Materials and Application Technology, Ningbo Institute of Material Technology and Engineering, Chinese Academy of Sciences, Ningbo 315201, China}
\author{Y. H. Wu}
\affiliation{Department of Electrical and Computer Engineering, National University of Singapore, 4 Engineering Drive 3 117583, Singapore}
\author{S. Zhang}
\email{zhangshu@email.arizona.edu}
\affiliation{Department of Physics, University of Arizona, Tucson, Arizona 85721, USA}
\author{Run-Wei Li}
\email{runweili@nimte.ac.cn}
\affiliation{Key Laboratory of Magnetic Materials and Devices, Ningbo Institute of Material Technology and Engineering, Chinese Academy of Sciences, Ningbo, Zhejiang 315201, China}
\affiliation{Laboratory of Magnetic Materials and Application Technology, Ningbo Institute of Material Technology and Engineering, Chinese Academy of Sciences, Ningbo 315201, China}
\date{\today}

\begin{abstract}
We report an investigation of transverse Hall resistance and longitudinal resistance on Pt thin films sputtered on epitaxial LaCoO$_3$ (LCO) ferromagnetic insulator films. The LaCoO$_3$ films were deposited on several single crystalline substrates [LaAlO$_3$ (LAO), (La,Sr)(Al,Ta)O$_3$ (LSAT), and SrTiO$_3$ (STO)] with (001) orientation. The physical properties of LaCoO$_3$ films were characterized by the measurements of magnetic and transport properties. The LaCoO$_3$ films undergo a paramagnetic to ferromagnetic (FM) transition at Curie temperatures ranging from 40 K to 85 K, below which the Pt/LCO hybrids exhibit significant extraordinary Hall resistance (EHR) up to 50 m$\Omega$ and unconventional magnetoresistance (UCMR) ratio $\Delta$$\rho$/$\rho_0$ about $1.2 \times 10^{-4}$, accompanied by the conventional magnetoresistance (CMR). The observed spin transport properties share some common features as well as some unique characteristics when compared with well-studied Y$_3$Fe$_5$O$_{12}$-based Pt thin films. Our findings call for new theories since the extraordinary Hall resistance and magnetoresistance cannot be consistently explained by the existing theories.
\begin{description}
\item[PACS number(s)]
72.25.Mk, 75.70.-i, 75.47.-m, 75.76.+j
\end{description}
\end{abstract}

\maketitle
\section{\label{sec:level1}INTRODUCTION}

The interplay between spin and charge transport in nonmagnetic metal/ferromagnetic insulator (NM/FMI) hybrids gives rise to various interesting phenomena, such as spin injection~\cite{Ohno1999,Jedema2001}, spin pumping~\cite{Heinrich2011,Rezende2012,Kajiwara2010}, and spin Seebeck~\cite{Uchida2008,Uchida2010}. The previous investigations on NM/FMI hybrids, e.g., Pt/Y$_3$Fe$_5$O$_{12}$, Pt/CoFe$_2$O$_4$, Pd/Y$_3$Fe$_5$O$_{12}$, and Ta/Y$_3$Fe$_5$O$_{12}$, demonstrated a new-type of magnetoresistance (MR)~\cite{Miao2014,Althammer2013,Isasa2014,Lin2014,Hahn2013} in which the resistivity of the film, $\rho$, has an unconventional angular dependence, namely,
\begin{equation}
\rho = \rho_0 - \Delta \rho \left[\hat{\bf m} \cdot (\hat{\bf z}\times \hat{\bf j}) \right]^2
\end{equation}
where $\hat{\bf m}$ and $\hat{\bf j}$ are unit vectors in the directions of the magnetization and the current, respectively,  and $\hat{\bf z}$ represents the normal vector perpendicular to the plane of the layers. The above unusual angular dependent resistivity differs from the conventional magnetoresistance (CMR) in which $\rho = \rho_0 + \Delta \rho (\hat{\bf m} \cdot \hat{\bf j})^2$, and thus several theoretical models have been developed to explain the effect. One of the most successful models is spin Hall magnetoresistance (SMR) which is built on the spin Hall and inverse spin Hall effects~\cite{Hirsch1999,Wunderlich2005,Kato2004,Tatara2006,Kajiwara2006,Kimura2007,Nakayama2013,Chen2013}: an electric current (${\bf j}_e$) induces a spin current ${\bf j}_s = \theta_\textup{SH} {\bf j}_e \times \boldsymbol{\sigma}$, where $\theta_\textup{SH}$ is the spin Hall angle and $\boldsymbol{\sigma}$ is the spin of the conduction electron. Such induced spin current in turn generates an electric current whose direction is opposite to the original ${\bf j}_e$, and thus the spin Hall and inverse spin Hall effects increase the resistivity by a factor of $1+\theta_\textup{SH}^2$. However, in the case of a thin film in contact with a magnetic insulator, the spin current may be either reflected back, which would reduce the spin current in the film, or absorbed via spin transfer torques, which would preserve the spin current in the film~\cite{Nakayama2013,Chen2013}. The reflection is strongest when the magnetization direction of the insulator is parallel to the spin current, therefore the spin current is least when $\hat{\bf m}$ is parallel to $\hat{\bf z}\times \hat{\bf j}$, leading to the minimum resistivity as described in Eq.~(1). Another model is based on the spin-dependent scattering and Rashba spin-orbit coupling (SOC)~\cite{Zhang2014-2,Grigoryan2014}. When the electron scatters off the magnetic interface, the resistivity for spin up and down relative to the direction of the magnetization is different. In the presence of interface Rashba spin-orbit interaction, two spin channels are mixed and an additional resistance appears. It has been found that such mechanism also gives rise to an unconventional magnetoresitance (UCMR) and an extraordinary Hall resistance (EHR)~\cite{Zhang2014-2,Grigoryan2014}.

\begin{figure}[tbp]
     \begin{center}
     \includegraphics[width=3.4in,keepaspectratio]{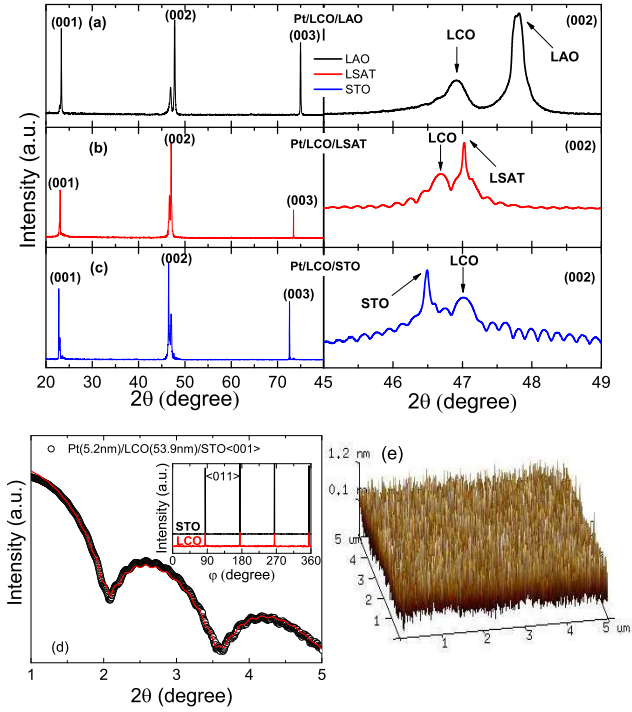}
     \end{center}
     \caption{(Color online) Three representative 2$\theta$-$\omega$ XRD patterns for Pt/LCO/LAO (a), Pt/LCO/LSAT (b), and Pt/LCO/STO (c) hybrids. The enlarged plot of (002) reflections are plotted in the right panel of (a)-(c). (d) The XRR spectrum of a representative Pt/LCO hybrid. The solid red line is a fit to the experimental data. The inset shows the $\varphi$-scan of Pt/LCO/STO hybrid. (e) Three dimensional plot of the AFM image for Pt/LCO/STO hybrid.}
     \label{fig1}
\end{figure}

However, the other spin transport properties are not obviously supporting the SMR picture. The EHR has shown rich characteristics such as unusual temperature dependence of Hall conductivity whose magnitude and sign are highly non-trivial~\cite{Miao2014,Huang2012}. If one were to apply the SMR model, one would have to use an unphysical imaginary part of the spin mixing conductance parameter. In particular, it requires an arbitrary temperature dependent mixing parameter to qualitatively fit the EHR data. Furthermore, the recent magnetoresistance data at high magnetic field reveals that the UCMR in the form of Eq.~(1) persists even after the magnetization is saturated~\cite{Miao2014}. Such high field UCMR and unusual EHR data indicate that the transport properties in NM/FMI hybrids are far from understood.

Up to now, most of the UCMR has been reported in the Y$_3$Fe$_5$O$_{12}$ (YIG)-based NM/FMI hybrids~\cite{Miao2014,Althammer2013,Isasa2014,Lin2014,Hahn2013}. The large difference of magnitude of UCMR observed in Pt/YIG indicates the importance of interface quality~\cite{Althammer2013}. In order to clarify the nature of the UCMR and EHR, it is highly desirable to investigate other NM/FMI hybrids. Ferromagnetic (FM) transition below $T_\textup{C} \approx$ 85 K has been recently observed in a perovskite-type LaCoO$_3$ (LCO) epitaxial film~\cite{Fuchs2007,Fuchs2008,Herklotz2009,Mehta2015}, and its FM insulating ground state has been theoretically proposed and experimentally observed~\cite{Freeland2008,Hsu2012}. In contrast to the YIG film, which exhibits extreme high Curie temperature ($T_\textup{C} \approx$ 540 K)~\cite{Uchida2015}, it is much easier to investigate the difference of spin transport properties between the paramagnetic and ferromagnetic states in LCO-based hybrids. Moreover, the LCO exhibits much simper crystal structure and can be deposited on various single crystalline substrates, and its Curie temperature can be tuned by epitaxial strain~\cite{Fuchs2008}. We first found that the UCMR and EHR disappear in the paramagnetic state of LCO insulating films. In this paper, the temperature, magnetic field and angular dependence of transverse Hall resistance R$_{xy}$ and longitudinal resistance R$_{xx}$ were investigated in Pt/LCO hybrids with various Pt thicknesses. The paper is organized as follows: After the introduction, we describe the sample growth and detailed experimental procedures in Sec. II. Section III(A) characterizes the structure information of Pt/LCO hybrids and Sec. III (B) describes the magnetic and transport properties of LCO films deposited on different single crystalline substrates. In Secs. III (C) and III (D), we present the experimental results of transverse Hall resistance R$_{xy}$ and longitudinal resistance R$_{xx}$. Finally, discussion and summary are given in Sec. III (E) and Sec. IV, respectively.

\begin{figure}[tbp]
\begin{center}
\includegraphics[width=3.4in,keepaspectratio]{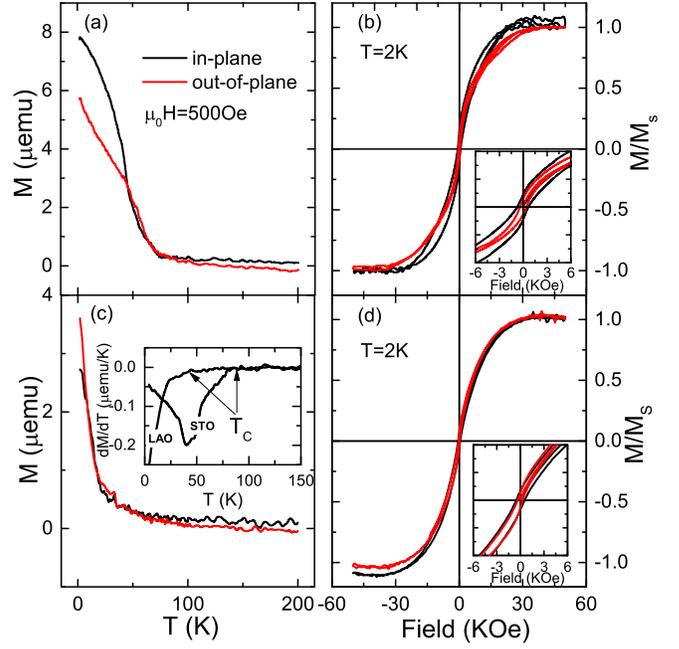}
\end{center}
\caption{(Color online) Temperature dependence of field-cooled magnetization for LCO/STO film (a) and LCO/LAO film (c). The inset of (c) plots the derivative of in-plane magnetization with respect to temperature d$M$/d$T$. (b) and (d) show the field dependence of magnetization normalized to the saturated magnetization $M_\textup{s}$ for LCO/STO and LCO/LAO films measured at $T$ = 2 K, respectively; the low insets expand the low field regions. For the in-plane (out-of-plane) magnetization, the magnetic field  was applied parallel (perpendicular) to the film surface.}
\label{fig2}
\end{figure}

\section{EXPERIMENTAL DETAILS}

The Pt/LCO hybrids were prepared in a combined ultra-high vacuum (10$^{-9}$ Torr) pulse laser deposition (PLD) and magnetron sputter system. The high quality LCO films with a thickness of approximately 50 nm were epitaxially grown on various (001)-orientated single crystalline substrates via PLD technique, i.e., LaAlO$_3$ (LAO), (La,Sr)(Al,Ta)O$_3$ (LSAT), and SrTiO$_3$ (STO). The stoichiometric sinter LCO target was used for epitaxial deposition. The deposition temperature and the oxygen background pressure were kept at 750 $^\circ$C and 50 mTorr, respectively. After deposition, the films were annealed at this deposition condition for one hour to ensure a complete and homogeneous oxygenation. The polycrystalline Pt films were sputtered at room temperature in 4 mTorr argon atmosphere in an in situ process with the Pt thickness in a range of 2 nm $\leq$ t$_\textup{Pt}$ $\leq$ 15 nm. All films were patterned into Hall bar geometry (central area: 0.3 mm $\times$ 10 mm and electrode: 0.3 mm $\times$ 1 mm). The thickness and crystal structure of films were characterized by using Bruker D8 Discover high-resolution x-ray diffractometer (HRXRD). The thickness was estimated by using the software package LEPTOS (Bruker AXS). The surface topography of the films was measured in Bruker icon atomic force microscope (AFM). The magnetic properties of the films were studied by using quantum-design magnetic properties measurement system (SQUID VSM-7 T). The measurements of transverse Hall resistance R$_{xy}$ and longitudinal resistance R$_{xx}$ were carried out in quantum-design physical properties measurement system with a rotating state (PPMS-9 T) at a temperature range of 2-300 K.

\section{RESULTS AND DISCUSSION}
\subsection{Structural characterization}
Figures 1(a)-(c) plot representative room temperature 2$\theta$-$\omega$ XRD scans of LCO films deposited on (001)-oriented LAO (top), LSAT (middle) and STO (bottom) single crystalline substrates, with other films showing similar patterns. No indication of impurities or misorientation was detected in the full range of 2$\theta$-$\omega$ scan (10-80 degree). The enlarged plot of (002) reflections are presented in the right panel of Figs. 1(a)-(c), where the arrows mark the reflection peak of LCO films. Clear Lauer oscillations indicate high quality and uniformity of epitaxial LCO films on LSAT and STO substrates. Whereas, no oscillation can be found on LAO substrate. One possible reason is that the mismatch of lattice parameters between LCO film and substrate is much larger on LAO substrate than on LSAT or STO substrate. The difference of 2$\theta$ value $\Delta_{2\theta}$ ($\Delta_{2\theta}$ = $2\theta_\textup{LCO}$ - $2\theta_\textup{LAO}$) between LCO and LAO (004)-reflection peak is -0.868 degree, while for LSAT and STO substrates, the $\Delta_{2\theta}$ value are -0.344 and +0.526 degree, respectively. The film thickness was determined by fitting the x-ray reflectivity (XRR) spectra. For instance, we plot XRR spectrum of Pt(5.2 nm)/LCO(53.9 nm)/STO hybrid in Fig. 1(d). It is noted that the actual Pt thicknesses derived from the simulations of XRR spectra are almost identical to the nominal thicknesses. For clarity, the nominal Pt thicknesses rather than actual thicknesses are used in this paper. The epitaxial nature of LCO films was characterized by $\varphi$-scan measurement with a fixed 2$\theta$ value at the (011) reflection of substrate and LCO film. For instance, the $\varphi$-scans of Pt/LCO/STO hybrid are plotted in the inset of Fig. 1(d), with the other hybrids showing similar patterns. The atomic force microscope surface topographies of Pt/LCO/LAO (PLL), Pt/LCO/LSAT (PLLA), and Pt/LCO/STO (PLS) hybrids reveal the surface roughness ranging from 0.2 nm to 0.6 nm, indicating atomical flat of prepared films. Figure 1(e) plot a representative three dimensional AFM topography of PLS hybrid.

\begin{figure}[tbp]
\begin{center}
\includegraphics[width=3.4in,keepaspectratio]{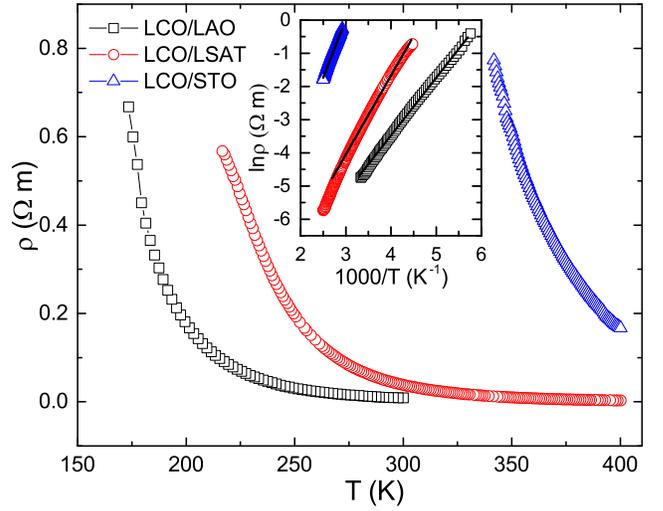}
\end{center}
\caption{(Color online) Temperature dependence of the electrical resistivity of LCO films. The thickness of LCO films is about 50 nm. The electrical resistivity was measured down to 2 K. The inset plots the logarithmic electrical resistivity ln$\rho$ versus 1/$T$. The solid black lines are fits to $\rho$ = $\rho_0$ e$^{\varepsilon/\textup{k}_BT}$.}
\label{fig3}
\end{figure}

\subsection{Physical properties of LaCoO$_3$ films}

\begin{figure}[tbp]
     \begin{center}
     \includegraphics[width=3.4in,keepaspectratio]{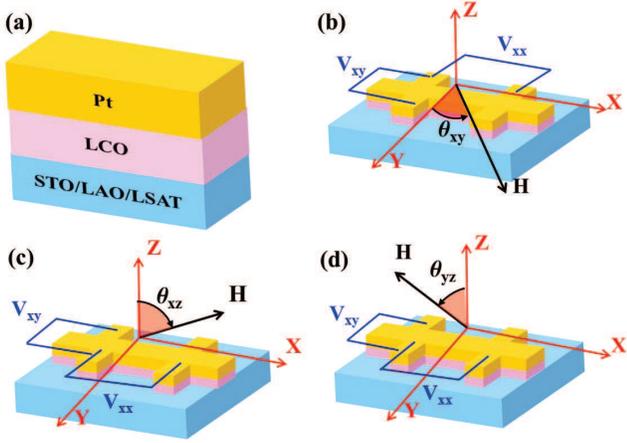}
     \end{center}
     \caption{(Color online) (a) Structure of prepared thin films. (b)-(c) Schematic plot of longitudinal resistance (R$_{xx}$) and transverse Hall resistance (R$_{xy}$) measurements and notations of different field scans in the patterned Hall bar Pt/LCO hybrids. The magnetic field can be applied in the $xy$, $xz$, and $yz$ planes with angles $\theta_{xy}$, $\theta_{xz}$, and $\theta_{yz}$ relative to the $y$-, $z$-, and $z$-axis.}
     \label{fig4}
\end{figure}

\begin{figure}[tbp]
     \begin{center}
     \includegraphics[width=3.4in,keepaspectratio]{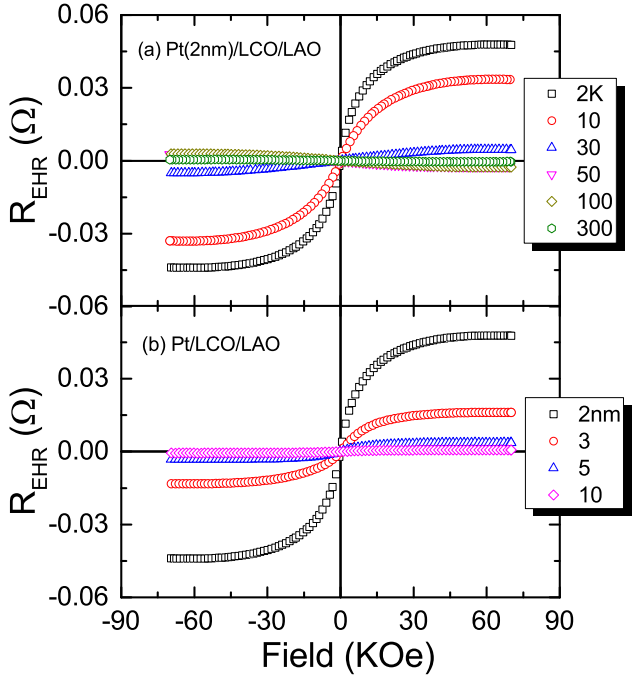}
     \end{center}
     \caption{(Color online) (a) The extraordinary Hall resistance R$_\textup{EHR}$ of 2 nm Pt thin film sputtered on LCO/LAO film at different temperatures down to 2 K. (b) The R$_\textup{EHR}$ of Pt/LCO/LAO hybrids with different Pt thickness, i.e., 2, 3, 5, and 10 nm, which were measured at 2 K. The R$_\textup{EHR}$ can be derived by subtracting the linear background of ordinary Hall resistance.}
     \label{fig5}
\end{figure}

According to the previous investigations, the magnetic properties and insulating nature of LaCoO$_3$ film can be tuned by epitaxial strain on different single crystalline substrates~\cite{Fuchs2007,Fuchs2008,Herklotz2009,Mehta2015}. We have carefully optimized the depositing conditions by measuring the magnetic and transport properties to get ferromagnetic insulating LCO films. The ferromagnetic ordering temperatures of LCO epitaxial films are reproducible. The dc magnetization of LCO films were measured as functions of temperature and magnetic field. Figures 2(a) shows the temperature dependence of the magnetization for LCO/STO film, which was measured in a magnetic field of $\mu_0$$H$ = 500 Oe applied both parallel to the film surface (in-plane, black line) and perpendicular to the film surface (out-of-plane, red line). Both in-plane and out-of-plane magnetization exhibit a sharp increase at $T_\textup{C} \approx 85$ K, indicating that the LCO/STO film undergoes a FM transition at Curie temperature $T_\textup{C}$, whose value can be obtained from the d$M$/d$T$ [see inset of Fig. 2(c)]. The ferromagnetic ground state can be further corroborated by a hysteresis loop in $M(H)$. As shown in Fig. 2(b), both in-plane and out-of-plane magnetization show an obvious saturated hysteresis loop at $T = 2$ K, with the  saturation field reaching 30 KOe. Magnetization of LCO film on LAO substrate, measured in a magnetic field of $\mu_0$$H$ = 500 Oe, are presented in Fig. 2(c). In contrast to a sharp FM transition in LCO/STO film, both in-plane and out-of-plane magnetization of LCO/LAO film show a broad transition around $T_\textup{C} \approx 40$ K, as the arrow indicated in the inset of Fig. 2(c). The hysteresis loops at $T = 2$ K with saturation field of 30 KOe shown in Fig. 2(d) also confirm a FM state in LCO/LAO film. Finally, the LCO film on LSAT substrate exhibits similar behaviors to the LCO/LAO film with a Curie temperature of 75 K (not show here). It is noted that the in-plane lattice parameter of LAO, LSAT, and STO substrates are 3.810${\AA}$, 3.868${\AA}$, and 3.905${\AA}$, respectively. Thus, the Curie temperatures $T_\textup{C}$ of LCO film increases with the in-plane film lattice parameter, consistent with previous results~\cite{Fuchs2008}. As shown in the insets of Fig. 2(b) and (d), all LCO films have similar low coercivity and remanence, indicating weak magnetic anisotropy in the epitaxial LCO films.

In order to check the insulating nature of deposited LCO films, we also carried out the measurements of electrical resistivity as a function of temperature. As shown in Fig. 3, though the LCO/STO exhibits a sharp FM transition, the electrical resistivity displays an extreme insulating state below 350 K, where the electrical resistivity is too large to be measured in the PPMS systems. In comparison, the electrical resistivity of LCO/LAO and LCO/LSAT films are much smaller than LCO/STO film, but they also exceed the measurement limit below 170 K and 220 K, respectively. Above these measurable temperatures, the electrical resistivity of LCO films can be formulated by $\rho$ = $\rho_0$ e$^{\varepsilon/\textup{k}_BT}$ (see solid black lines in the inset of Fig. 3), where $\varepsilon$ and k$_\textup{B}$ are activation energy and Boltzmann constant. The estimated energy gaps are $\varepsilon$ = 149.87 meV, 205.36 meV, and 300.04 meV for LCO/LAO, LCO/LSAT, and LCO/STO, respectively. Thus, the transport properties of Pt/LCO hybrids are only associated with Pt films below there temperatures.

\begin{figure}[tbp]
     \begin{center}
     \includegraphics[width=3.2in,keepaspectratio]{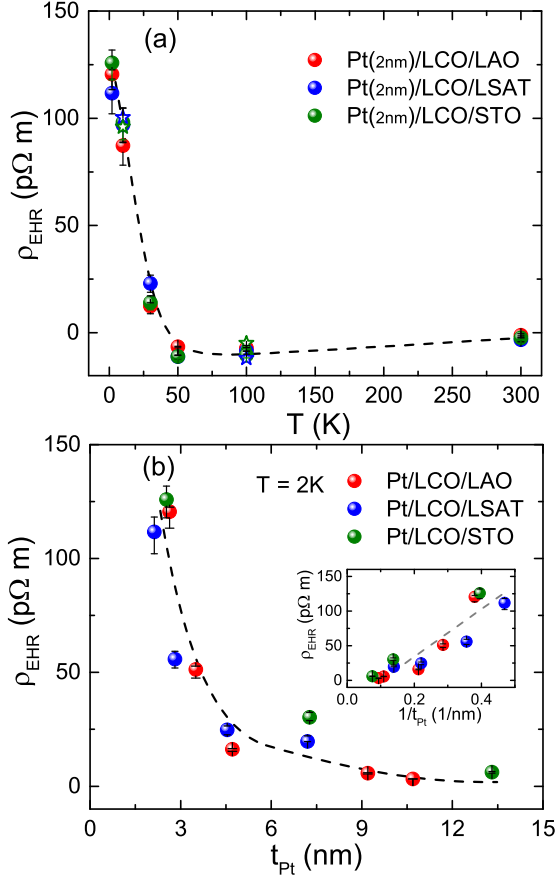}
     \end{center}
     \caption{(Color online) (a) The R$_\textup{EHR}$ as a function of temperature for PLL2, PLLA2, and PLS2 hybrids. (b) Thickness dependence of R$_\textup{EHR}$ for PLL, PLLA, and PLS hybrids at 2 K. The inset of (b) plots the R$_\textup{EHR}$ versus 1/t$_\textup{Pt}$. All R$_\textup{EHR}$ are averaged by [R$_\textup{EHR}$(7 T)- R$_\textup{EHR}$(-7 T)]/2. The “star” symbols represent the data derived from the fitting results in Fig. 7(b). The dashed lines are guide to the eyes. The error bars are a result of subtracting ordinary Hall resistance in different field ranges.}
     \label{fig6}
\end{figure}

\subsection{Transverse Hall resistance of Pt/LCO hybrids}

As shown in Fig. 4, all the Pt/LCO hybrids were patterned into Hall bar geometry and the electric current is applied along the $x$-axis. The $\theta_{xy}$, $\theta_{xz}$, and $\theta_{yz}$ are defined as angles between the applied magnetic field and the electric current. The $\theta_{xy}$ scan accesses the longitudinal ($\rho_\parallel$, $H \parallel$ $x$) and the transverse resistivity ($\rho_\textup{T}$, $H \parallel$ $y$), while the $\theta_{xz}$ and $\theta_{yz}$ scans can reach the perpendicular resistivity ($\rho_\perp$, $H \parallel$ $z$).

\begin{figure}[tbp]
     \begin{center}
     \includegraphics[width=3.0in,keepaspectratio]{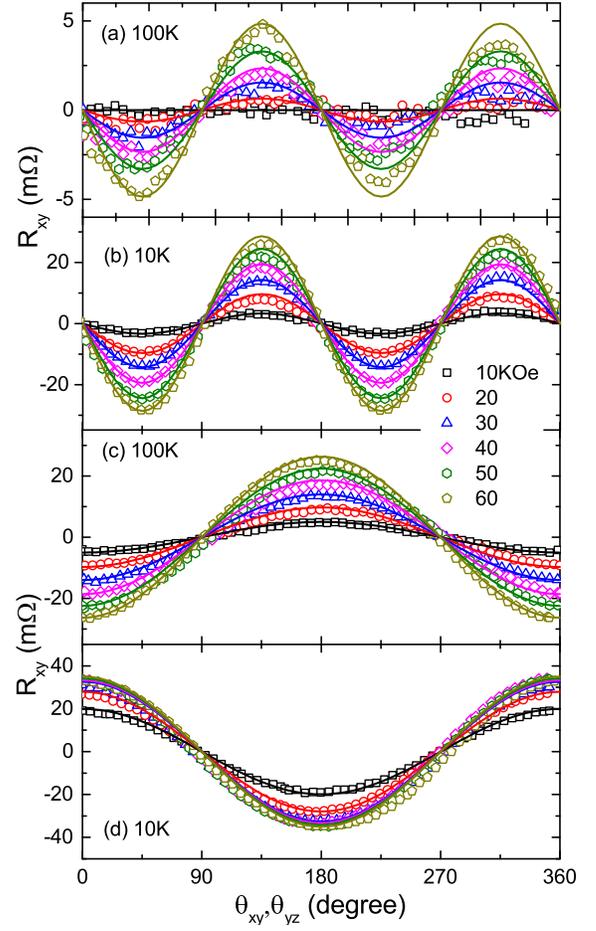}
     \end{center}
     \caption{(Color online) Planar Hall resistance under various magnetic fields at temperatures of 100 K (a) and 10 K (b) for Pt(2 nm)/LCO/LAO hybrid. The solids lines through the data are fits to R$_\textup{xy}$ = R$_1$sin$\theta$cos$\theta$. Angular dependence of the Hall resistance R$_\textup{xy}$ under various magnetic fields at temperatures of 100 K (c) and 10 K (d) for Pt(2 nm)/LCO/LAO hybrid; The magnetic field sweeps within $yz$-plane.}
     \label{fig7}
\end{figure}

In this section, we discuss the results of transverse Hall resistance R$_\textup{xy}$ of Pt/LCO hybrids with perpendicular magnetic field ranging from -70 KOe to 70 KOe and temperature ranging from 2 K to 300 K. In Pt thin film, the ordinary hall resistance (OHR) R$_\textup{OHR}$, which is proportional to the external field, is subtracted from the measured  R$_\textup{xy}$, i.e., R$_\textup{EHR}$  = R$_{xy}$ - R$_\textup{OHR}$$\times \mu_0H$,  R$_\textup{EHR}$ is extraordinary Hall resistance. In a ferromagnetic metal, R$_\textup{EHR}$ is proportional to the out-of-plane magnetization. The resulting R$_\textup{EHR}$ as a function of magnetic field for Pt(2 nm)/LCO/LAO (PLL2) hybrid are exhibited in Fig. 5(a) in a temperature range of 2-300 K, with the Pt(2 nm)/LCO/LSAT (PLLA2) and Pt(2 nm)/LCO/STO (PLS2) hybrids showing similar behaviors. R$_\textup{EHR}$ saturates for $\mu_0H > 30$ KOe, consistent with the $M(H)$ results in Fig. 2. Upon decreasing the temperature, the PLL2 shows obvious R$_\textup{EHR}$ below the Curie temperature of LCO film. For examples, at $T$ = 2 K, the saturated R$_\textup{EHR}$ reaches 45.8 m$\Omega$, 52.4 m$\Omega$, and 49.8 m$\Omega$ for PLL2, PLLA2, and PLS2 hybrids, respectively. However, at temperatures above the Curie temperature of LCO film, e.g., $T$ = 100 K, no sizable R$_\textup{EHR}$ can be found for all hybrids, implying intimate relationship between R$_\textup{EHR}$ and LCO ferromagnetism. The temperature dependence of the derived R$_\textup{EHR}$ for PLL2, PLLA2, and PLS2 hybrids are summarized in Fig. 6 (a). As can be seen, the R$_\textup{EHR}$ is significantly temperature dependent. Below the Curie temperature of LCO film, the R$_\textup{EHR}$ increases sharply as decreasing the temperature and the R$_\textup{EHR}$ changes its sign below 50 K, similar behaviors were previously reported in Pt/YIG hybrids~\cite{Miao2014,Huang2012}. We also plot the R$_\textup{EHR}$ of PLL hybrids with different Pt thickness at $T$ = 2 K in Fig. 5(b). As the Pt thickness increased, the R$_\textup{EHR}$ dramatically decreases, and becomes almost negligible for t$_\textup{Pt} >$ 10 nm. For example, for Pt(15 nm)/LCO/STO hybrid, the R$_\textup{EHR}$ is about 0.46 m$\Omega$ at 2 K, which is two orders of magnitude less than PLS2 hybrid. The thickness dependence of the R$_\textup{EHR}$ are summarized in Fig. 6(b). The R$_\textup{EHR}$ scales as 1/t$_\textup{Pt}$ [see inset of Fig. 6(b)], indicating interfacial contributions dominate the R$_\textup{EHR}$. Similar behaviors were also reported previously in Pd/YIG hybrid~\cite{Lin2013}.

\begin{figure}[tbp]
     \begin{center}
     \includegraphics[width=3.2in,keepaspectratio]{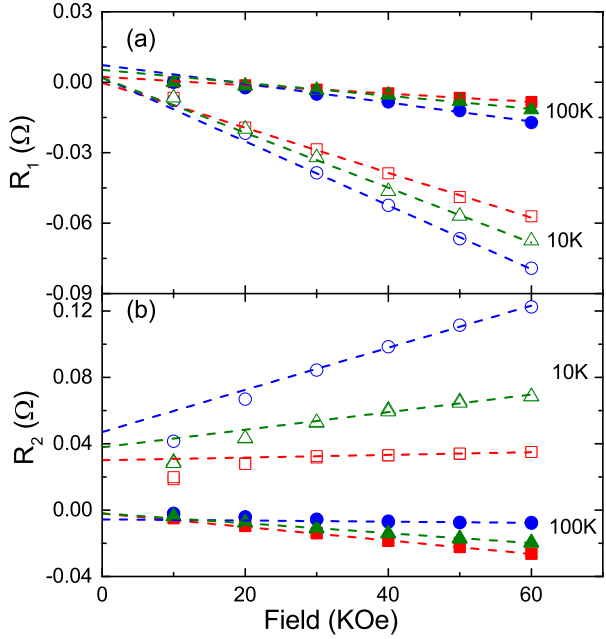}
     \end{center}
     \caption{(Color online) Fitting parameters of equations discussed in the context. Different symbols represent different hybrids: PLL (square), PLLA (triangle), and PLS (circle). The dashed lines represent linear fits to the magnetic field. The solid symbols represent the data at 100 K, while the open ones represent the 10 K data.}
     \label{fig8}
\end{figure}

Next, we carried out the measurements of angular dependence of R$_\textup{xy}$ at temperatures of 100 K and 10 K. For clarity, only the data of PLL2 are presented, with the PLLA2 and PLS2 showing similar behaviors. When the magnetization is in the plane, the angular dependence of R$_\textup{xx}$ and R$_\textup{xy}$ take the following formulas~\cite{McGuire1975}:
\begin{equation}
R_\textup{xx} \propto \rho_\textup{T} +(\rho_\parallel - \rho_\textup{T})\textup{cos}^2\theta
\end{equation}
\begin{equation}
R_\textup{xy} \propto (\rho_\parallel - \rho_\textup{T})\textup{sin} \theta \textup{cos}\theta
\end{equation}
Equation (2) is known as anisotropic magnetoresistance (AMR) and Eq. (3) is known as planar Hall effect (PHE). Figures 7(a)-(b) plot the angular dependence of $R_\textup{xy}$ in magnetic fields up to 60 KOe. As the solid lines shown, R$_\textup{xy}$ can be well described by R$_1$sin$\theta$cos$\theta$, where R$_1$ is proportional to the difference between parallel and transverse resistance. The derived magnitude R$_1$ for all hybrids, including PLL2, PLLA2, and PLS2, are summarized as a function of magnetic field in Fig. 8 (a). For $T$ = 10 K, which is far below the Curie temperature of LCO film, R$_1$ is almost 7 times larger than that of 100 K. Such enhanced R$_1$ value is likely caused by the induced interfacial moments due to the magnetic proximity effect (MPE) at the Pt/LCO interface at low temperatures. It is noted that, at $T$ = 10 K, for $\mu_0$H $>$ 30 KOe, R$_1$ is linear in field, which is consistent with saturated field in the magnetization (see details in Fig. 2).

\begin{figure}[tbp]
     \begin{center}
     \includegraphics[width=3.2in,keepaspectratio]{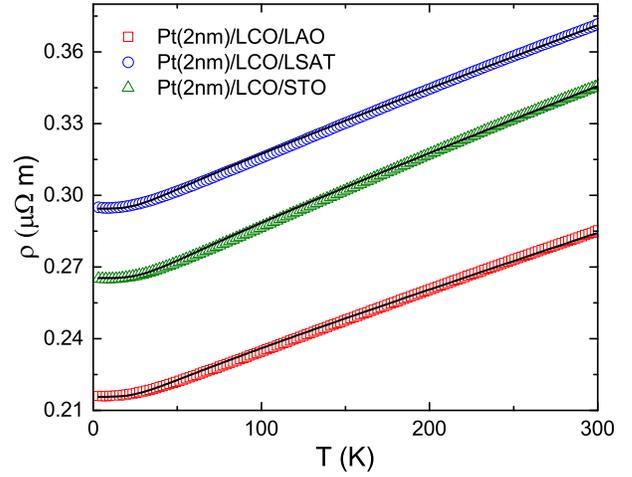}
     \end{center}
     \caption{(Color online)  Temperature dependence of the electrical resistivity for PLL2, PLLA2, and PLS2 hybrids. The solid lines are fits to Eq. (4) in the text.}
     \label{fig9}
\end{figure}

The angular dependence of transverse Hall resistance with the magnetic field rotating within $xz$- and $yz$-plane were also measured. R$_\textup{xy}$ as a a function of angle $\theta_{yz}$ for various magnetic field up to 60 KOe are shown in Figs. 7(c)-(d), with R$_\textup{xy}$($\theta_{xz}$) showing similar behaviors. As the solid lines shown, R$_\textup{xy}$($\theta_{yz}$) at different magnetic fields can be well described by R$_\textup{xy}$ = R$_2$cos$\theta$. In contrast to Pt/YIG, no additional higher than linear order contributions (cos$^3$$\theta$) was observed in Pt/LCO hybrids~\cite{Meyer2015}. The field dependence of R$_2$ for all hybrids, including PLL2, PLLA2, and PLS2, are summarized in Fig. 8 (b). It can be clearly seen that the R$_2$ changes its sign as the temperature decreases, similar to the results in Fig. 5. The EHR contribution to R$_\textup{xy}$ can be separated from OHR contribution by fitting the data to R$_2$ = R$_\textup{EHR}$ + R$_\textup{OHR}$$\times$$\mu_0H$. The derived R$_\textup{EHR}$ are also shown in Fig. 6(a) (see open star symbols) as a function of temperature, which are consistent with the field dependence measurements of R$_\textup{xy}$.

\subsection{Longitudinal resistance of Pt/LCO hybrids}

\begin{figure}[tbp]
     \begin{center}
     \includegraphics[width=3.2in,keepaspectratio]{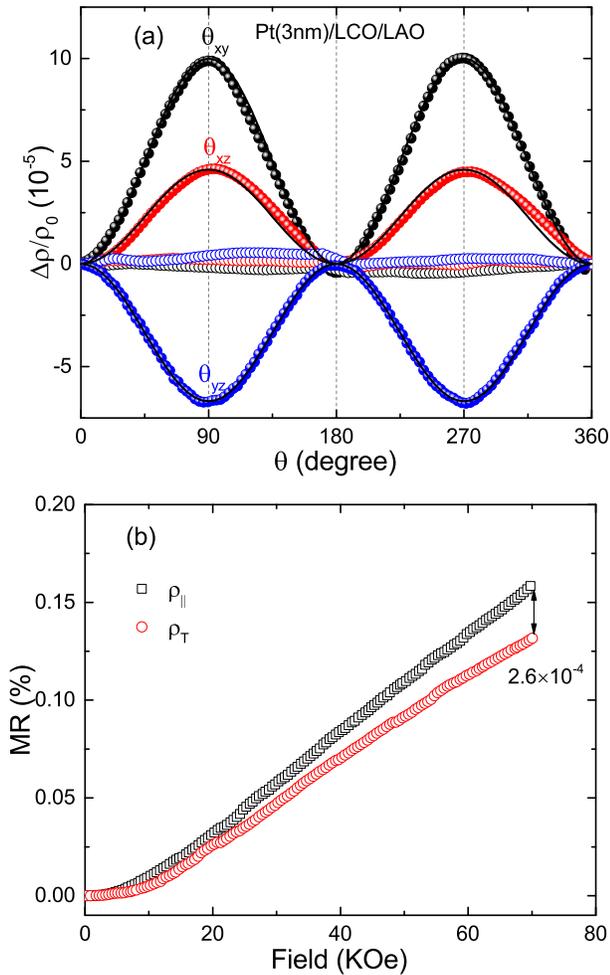}
     \end{center}
     \caption{(Color online) (a) Angular dependence of the MR for Pt (3 nm)/LCO/LAO hybrid for $\theta_{xy}$, $\theta_{xz}$, and $\theta_{yz}$ scans. The data were measured at 10 K (solid symbols) and 100 K (open symbols) in a field of $\mu_0$H = 40 KOe. The solid lines through the data are fits to (cosine)$^2$ with 90 degree phase shift. (b) The field dependence of longitudinal ($\rho_\parallel$) and transverse ($\rho_\textup{T}$) MR ratio for Pt (3 nm)/LCO/LAO hybrid measured at $T$ = 2 K.}
     \label{fig10}
\end{figure}

\begin{figure}[tbp]
     \begin{center}
     \includegraphics[width=3.2in,keepaspectratio]{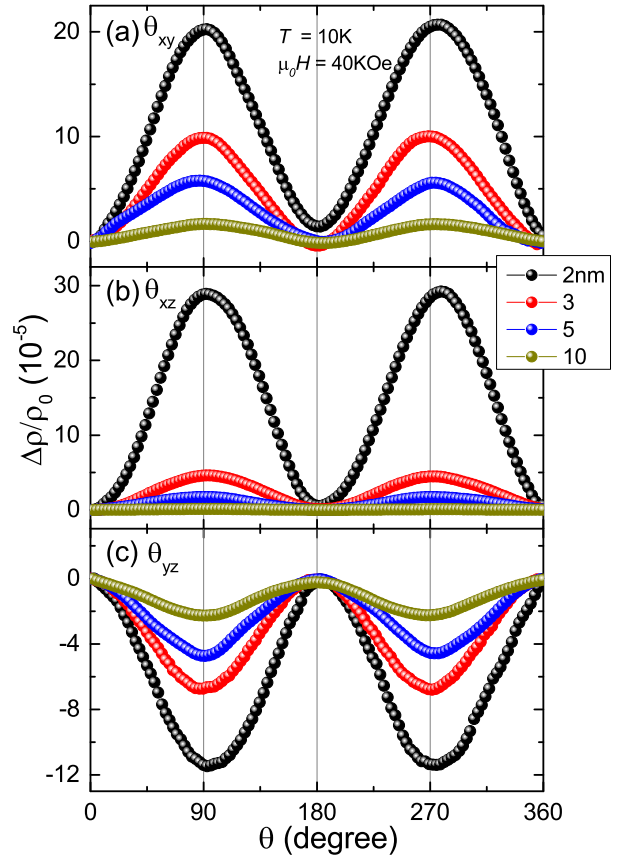}
     \end{center}
     \caption{(Color online) Angular dependence of MR for Pt/LCO/LAO hybrid with different Pt thickness for $\theta_{xy}$ (a), $\theta_{xz}$ (b), and $\theta_{yz}$ (c) scans. The data were measured at 10 K in a field of $\mu_0$H = 40 KOe, which is sufficiently strong to saturate the LCO magnetization.}
     \label{fig11}
\end{figure}

Figure 9 shows temperature dependence of the electrical resistivity for PLL2, PLLA2, and PLS2 hybrids. The electrical resistivity in all films increases linearly with temperature, exhibiting typical metallic behavior. The temperature dependence of the electrical resistivity can be well described by the Bloch-Gr\"{u}neisen relation~\cite{Meaden1969}:
\begin{equation}
\rho(T) = \rho_R + \frac{4R}{\Theta_R}(\frac{T}{\Theta_R})^5\int_0^{\frac{\Theta_R}{T}}\frac{x^5}{(e^x-1)(1-e^{-x})}dx
\end{equation}
where $\rho_R$ is the residual resistivity due to the static defects in the crystal lattice; second term represents the contribution of electron-phonon scattering, in which $\Theta_R$ is the Debye temperature and $R$ is the electron-phonon coupling constant. The solid lines through the data in Fig. 9 are fits to Eq. (4) and the fitting parameters are summarized in Table. I. The Debye temperatures $\Theta_R$ from the fit are much smaller than the value derived from electrical resistivity ($\Theta_R$ $\approx$ 240 K) and specific heat ($\Theta_D$ $\approx$ 225 K) of bulk Pt metal~\cite{Meaden1969,tari2003}, again indicating the interfacial scattering contributes mostly to the electrical resistivity.

\begin{table}[tbp]
\caption{The values of $\rho_R$, $R$, and $\Theta_R$ derived from fittings to Eq. (4).}
\label{tab:table1}\centering
\begin{ruledtabular}
\begin{tabular}{lcccc}
\textrm{hybrid}&
\textrm{$\rho_R$($\mu$$\Omega$m)}&
\textrm{$R$($\mu$$\Omega$mK$^{-1}$)}&
\textrm{$\Theta_R$(K)}\\
\colrule
PLL2 & 0.215 & 5.76 & 157.5\\
PLLA2 & 0.294 & 6.08 & 153.0\\
PLS2 & 0.265 & 8.46 & 176.0 \\
\end{tabular}
\end{ruledtabular}
\end{table}

We also measure the angular and field dependence of MR for PLL and PLS hybrids. Both hybrids exhibit similar behaviors, for clarity, only the results of PLL hybrids are presented in Figs. 10 and 11. Figure 10 (a) plots the angular dependence of MR for Pt(3 nm)/LCO/LAO (PLL3) hybrid at 10 K and 100 K. The anisotropic magnetoresistance is defined as $\Delta$$\rho$/$\rho_0$ = [$\rho$(M $\parallel$ I) - $\rho$(M $\perp$ I)]/$\rho_0$, $\rho_0$ is zero field resistivity. According to Eq. (1), when magnetic field scans within $xy$-plane, both CMR and UCMR contribute to the total AMR, and it is difficult to separate these two contributions from each other; for $xz$-plane, the magnetic field is always perpendicular to the $\bf m$$_y$, $\bf m$$_y$ is the $y$ component of magnetization unit vector, the UCMR should remain constant, and any resistance change can be attributed to CMR; for $yz$-plane, the electrical current is always perpendicular to the magnetization, the CMR should remain constant, and only UCMR can be achieved. The PLL3 hybrid demonstrates clear UCMR below the Curie temperature of LCO film, with the amplitude exceeding 7 $\times$ 10$^{-5}$ [see $\theta_{yz}$ scan in Fig. 10(a)]; the $\theta_{xz}$ scans also show obvious CMR, indicating the existence of induced ferromagnetism at the Pt/LCO interface. Similar behaviors were also observed in Pd/YIG hybrid, where the amplitude of $\theta_{xz}$ scan increases as decreasing the temperature, as expected from the MPE induced ferromagnetism~\cite{Lin2014}. However, at the temperatures above the $T_\textup{C}$ of LCO film, no MR oscillation can be found for all three field scans, and the resistance is almost independent of angle $\theta$, implying that the AMR observed in Pt/LCO hybrids is entangled with the ferromagnetism of LCO film. As shown in Fig. 10(b), the field dependence of longitudinal ($\rho_\parallel$, $H \parallel$ $x$) and transverse ($\rho_\textup{T}$, $H \parallel$ $y$) MR for Pt (3 nm)/LCO/LAO hybrid measured at $T$ = 2 K are presented. Here the MR ratio is defined as MR = [$\rho$($H$) - $\rho_0$]/$\rho_0$. According to Eq. (1), the difference between longitudinal and transverse resistivity ($\Delta$$\rho$ = $\rho_\parallel$ - $\rho_\textup{T}$) depends on the relative angle between magnetization and current direction. So the $\Delta$$\rho$ should be a constant after LCO has been saturated for $\mu_0$H $>$ 30 KOe (see details in Fig. 2). However, the $\Delta$$\rho$ continuously increases as increasing the magnetic field, which is twice larger at 70 KOe (2.6 $\times$ 10$^{-4}$) than at 40 KOe (1.3 $\times$ 10$^{-4}$). Such unsaturated MR was also observed in Pt/YIG hybrid~\cite{Miao2014}, though it is believed to be caused by the MPE at Pt/YIG interface, but its nature is still far from understood. We also studied the angular dependence of MR for PLL hybrids with different Pt thickness. As shown in Fig. 11, where exhibits all $\theta_{xy}$, $\theta_{xz}$, and $\theta_{yz}$ scans at 10 K, the UCMR was observed in all hybrids with its magnitude $\Delta$$\rho$/$\rho_0$ reaching the maximum value of 1.2 $\times$ 10$^{-4}$. For all three scans, the magnitude $\Delta$$\rho$/$\rho_0$ monotonically decreases as increasing the Pt thickness, which is different from the case of Pt/YIG, where the $\Delta$$\rho$/$\rho_0$ shows a maximum around t$_\textup{Pt}$ $\sim$ $\lambda_\textup{Pt}$ ($\lambda_\textup{Pt}$ is spin diffusion of Pt)~\cite{Althammer2013}. For t$_\textup{Pt}$ $>$ 10 nm, there is no measurable oscillation for all three field scans.

\subsection{Discussion}
Based on the above experimental data, we conclude that the observed extraordinary Hall resistance and unconventional magnetoresistance in Pt thin films are entangled with the ferromagnetism of LCO insulator. However, these spin transport properties can not be generally explained by the existing theory. The recent theoretical model, namely, SMR has been proposed to explain the observed UCMR in Pt/YIG hybrid, which is built on the spin Hall and inverse spin Hall effects and involves a conversion between the charge and spin current~\cite{Nakayama2013,Chen2013}. However, the SMR model is unable to describe the unusual temperature dependence of EHR whose magnitude and sign are highly non-trivial and the unsaturated UCMR at high fields~\cite{Miao2014,Huang2012}. There are at least three contributions to the observed EHR in Pt/LCO hybrids: magnetic proximity effect, spin Hall based SMR, and spin-dependent interface scattering. The Pt is near the stoner ferromagnetic instability and could become magnetic when contacts with ferromagnetic materials, as shown experimentally by x-ray magnetic circular dichroism (XMCD) and theoretically by first principle calculation in Pt/YIG hybrids~\cite{Lu2013,Qu2013}. By injecting the Au layer, no measurable EHR can be detected for Pt(3 nm)/Au(6 nm)/YIG, indicating the important role of MPE~\cite{Miao2014}. However, our observed EHR which shows a significantly enhanced EHR at lower temperature cannot be simply explained by MPE since the XMCD measurements in Pt/YIG has revealed a slight decreased average Pt moment at room temperature  (0.054 $\mu_\textup{B}$) compared to low temperature (0.076 $\mu_\textup{B}$ at 10 K)~\cite{Lu2013}. The SMR model based on spin Hall effect predicts an anomalous Hall-like resistance whose magnitude is determined by the imaginary part of the spin mixing conductance~\cite{Chen2013}. However, the SMR model fails to explain the EHR behaviors in our Pt/LCO as well as previously reported Pt/YIG hybrids: an arbitrary temperature dependence of the imaginary part of the spin mixing conductance parameter is required to qualitatively describe the temperature dependent EHR data. Particularly, the sign reversal cannot even be qualitatively explained. Finally, spin-dependent scattering at the interface, combined with the conventional skew-scattering and side-jump mechanisms~\cite{Nagaosa2010}, can also give rise to EHR in our Pt/LCO hybrids. However, there is no existing quantitative theory to compare with our results. We conclude that non-trivial EHR observed in Pt/LCO and other similar hybrids demands further theoretical and experimental investigations in order to clarify the dominating mechanisms. On the hand, it is also interesting to investigate the impacts of possible spin state change in LCO film on the spin transport properties of Pt/LCO hybrids.

\section{CONCLUSION}
In summary, we carried out the measurements of transverse Hall resistance R$_{xy}$ and longitudinal resistance R$_{xx}$ on Pt/LCO hybrids. All three types of hybrids, including Pt/LCO/LAO, Pt/LCO/LSAT, and Pt/LCO/STO, exhibit extraordinary Hall resistance and unconventional magnetoresistance below the Curie temperature of LCO films. The amplitude of unconventional magnetoresistance and the extraordinary Hall resistance on Pt/LCO hybrids are comparable to Pd/YIG and Pt/YIG hybrids. However, the observed spin transport properties can not be consistently explained by the existing theories, further investigations are needed to clarify this issue.

\begin{acknowledgments}
This work is financially supported by the National Natural Science Foundation of China (Grant Nos. 11274321, 11404349, 11174302, 51502314). S. Zhang was partially supported by the National Science Foundation (Grant No. ECCS-1404542).
\end{acknowledgments}


\begin{thebibliography}{10}

\bibitem{Ohno1999} Y. Ohno, D. K. Young, B. Beschoten, F. Matsukura, H. Ohno, and D. D. Awschalom, Nature (London) \textbf{402}, 790 (1999).

\bibitem{Jedema2001} F. J. Jedema, A. T. Filip, and B. J. van Wees, Nature (London) \textbf{410}, 345 (2001).

\bibitem{Heinrich2011} B. Heinrich, C. Burrowes, E. Montoya, B. Kardasz, E. Girt, Y. Y. Song, Y. Y. Sun, and M. Z. Wu, Phys. Rev. Lett. \textbf{107}, 066604 (2011).

\bibitem{Rezende2012} S. M. Rezende, R. L. Rodr\'{\i}guez-Su\'{a}rez, M. M. Soares, L. H. Vilela-Leão, D. Ley Dom\'{\i}nguez, and A. Azevedo, Appl. Phys. Lett. \textbf{102} , 012402 (2013).

\bibitem{Kajiwara2010} Y. Kajiwara, K. Harii, S. Takahashi, J. Ohe, K. Uchida, M. Mizuguchi, H. Umezawa, H. Kawai, K. Ando, K.Takanashi, S. Maekawa, and E. Saitoh, Nature (London) \textbf{464}, 262 (2010).

\bibitem{Uchida2008} K. Uchida, S. Takahashi, K. Harii, J. Ieda, W. Koshibae, K. Ando, S. Maekawa, and E. Saitoh, Nature (London) \textbf{455}, 778 (2008).

\bibitem{Uchida2010} K. Uchida, J. Xiao,	H. Adachi, J. Ohe, S. Takahashi, J. Ieda, T. Ota, Y. Kajiwara, H. Umezawa, H. Kawai, G. E. W. Bauer, S. Maekawa, and E. Saitoh, Nat. Mater. \textbf{9}, 894 (2010).

\bibitem{Miao2014} B. F. Miao, S. Y. Huang, D. Qu, and C. L. Chien, Phys. Rev. Lett. \textbf{112}, 236601 (2014).

\bibitem{Althammer2013} M. Althammer, S. Meyer, H. Nakayama, M. Schreier, S. Altmannshofer, M. Weiler, H. Huebl, S. Gepr\"{a}gs, M. Opel, R. Gross, D. Meier, C. Klewe, T. Kuschel, J. M. Schmalhorst, G. Reiss, L. M. Shen, A. Gupta, Y. T. Chen, G. E. W. Bauer, E. Saitoh, and S. T. B. Goennenwein, Phys. Rev. B \textbf{87}, 224401 (2013).

\bibitem{Isasa2014} M. Isasa, A. B.Pinto, S. V\'{e}lez, F. Golmar, F. S\'{a}nchez, L. E. Hueso, J. Fontcuberta and F. Casanova, Appl. Phys. Lett. \textbf{105}, 142402 (2014).

\bibitem{Lin2014} T. Lin, C. Tang, H. M. Alyahayaei, and J. Shi, Phys. Rev. Lett. \textbf{113}, 037203 (2014)..

\bibitem{Hahn2013} C. Hahn, G. de Loubens, O. Klein, M. Viret, V. V. Naletov, and J. Ben Youssef, Phys. Rev. B \textbf{87}, 174417 (2013).

\bibitem{Hirsch1999} J. E. Hirsch, Phys. Rev. Lett. \textbf{83}, 1834 (1999).

\bibitem{Wunderlich2005} J. Wunderlich, B. Kaestner, J. Sinova, and T. Jungwirth, Phys. Rev. Lett. \textbf{94}, 047204 (2005).

\bibitem{Kato2004} Y. K. Kato, R. C. Myers, A. C. Gossard, and D. D. Awschalom, Science \textbf{306}, 1910 (2004).

\bibitem{Tatara2006} E. Saitoh, M. Ueda, H. Miyajima and G. Tatara, Appl. Phys. Lett. \textbf{88}, 182509 (2006).

\bibitem{Kajiwara2006} Y. Kajiwara, K. Harii, S. Takahashi, J. Ohe, K. Uchida, M. Mizuguchi, H. Umezawa, H. Kawai, K. Ando, K. Takanashi, S. Maekawa, and E. Saitoh, Nature (London) \textbf{442}, 176 (2006).

\bibitem{Kimura2007} T. Kimura, Y. Otani, T. Sato, S. Takahashi, and S. Maekawa, Phys. Rev. Lett. \textbf{98}, 156601 (2007).

\bibitem{Nakayama2013} H. Nakayama, M. Althammer, Y. T. Chen, K. Uchida, Y. Kajiwara, D. Kikuchi, T. Ohtani, S. Gepr\"{a}gs, M. Opel, S. Takahashi, R. Gross, G. E. W. Bauer, S. T. B. Goennenwein, and E. Saitoh, Phys. Rev. Lett. \textbf{110}, 206601 (2013).

\bibitem{Chen2013}Y. T. Chen, S. Takahashi, H. Nakayama, M. Althammer, S. T. B. Goennenwein, E. Saitoh, and G. E. W. Bauer, Phys. Rev. B \textbf{87}, 144411 (2013).

\bibitem{Zhang2014-2} S. S. -L. Zhang and S. Zhang, J. Appl. Phys. \textbf{115}, 17C703 (2014).


\bibitem{Grigoryan2014} V. L. Grigoryan, W. Guo, G. E. W. Bauer, and Jiang Xiao, Phys. Rev. B \textbf{90}, 161412(R) (2014).

\bibitem{Huang2012} S. Y. Huang, X. Fan, D. Qu, Y. P. Chen, W. G. Wang, J. Wu, T. Y. Chen, J. Q. Xiao, and C. L. Chien, Phys. Rev. Lett. \textbf{109}, 107204 (2012).

\bibitem{Fuchs2007} D. Fuchs, C. Pinta, T. Schwarz, P. Schweiss, P. Nagel, S. Schuppler, R. Schneider, M. Merz, G. Roth, and H. v. L\"{o}hneysen, Phys. Rev. B \textbf{75}, 144402 (2007).

\bibitem{Fuchs2008} D. Fuchs, E. Arac, C. Pinta, S. Schuppler, R. Schneider, and H. v. L\"{o}hneysen, Phys. Rev. B \textbf{77}, 014434 (2008).

\bibitem{Herklotz2009} A. Herklotz, A. D. Rata, L. Schultz, and K. D\"{o}rr, Phys. Rev. B \textbf{79}, 092409 (2009).

\bibitem{Mehta2015} V. V. Mehta, N. Biskup, C. Jenkins, E. Arenholz, M. Varela, and Y. Suzuki, Phys. Rev. B \textbf{91}, 144418 (2015).

\bibitem{Freeland2008} J. W. Freeland, J. X. Ma, and J. Shi, Appl. Phys. Lett. \textbf{93}, 212501 (2008).

\bibitem{Hsu2012} H. Hsu, P. Blaha, and R. M. Wentzcovitch, Phys. Rev. B \textbf{85}, 140404(R) (2012).

\bibitem{Uchida2015} K. Uchida, Z. Y. Qiu, T. Kikkawa, R. Lguchi, and E. Saitoh, Appl. Phys. Lett. \textbf{106}, 052405 (2015).

\bibitem{Lin2013} T. Lin, C. Tang, and J. Shi, Appl. Phys. Lett. \textbf{103}, 132407 (2013).

\bibitem{McGuire1975} T. R. McGuire and R. I. Potter, IEEE Trans. Magn. \textbf{11}, 1018 (1975).

\bibitem{Meyer2015} S. Meyer, R. Schlitz, S. Gepr\"{a}gs, M. Opel, H. Huebl, R. Gross, and S. T. B. Goennenwein, Appl. Phys. Lett. \textbf{106}, 132402 (2015).

\bibitem{Meaden1969} G. T. Meaden, \emph{Electrical resistance of metals}, (Heywood, 1969).

\bibitem{tari2003} A. Tari, \emph{The specific heat of matter at low temperatures}, (World Scientific, 2003).

\bibitem{Lu2013} Y. M. Lu, Y. Choi, C. M. Ortega, X. M. Cheng, J. W. Cai, S. Y. Huang, L. Sun, and C. L. Chien,  Phys. Rev. Lett. \textbf{110}, 147207 (2013).

\bibitem{Qu2013} D. Qu, S. Y. Huang, J. Hu, R. Q. Wu, and C. L. Chien, \textbf{110}, 067206 (2013).

\bibitem{Nagaosa2010} N. Nagaosa, J. Sinova, S. Onoda, A. H. MacDonald, and N. P. Ong, Rev. Mod. Phys. \textbf{82}, 1539 (2010) and reference therein.

\end{thebibliography}
\end{document}